\documentclass[twocolumn,showpacs,amsmath,amssymb]{revtex4}
\usepackage{graphicx}
\usepackage{epsfig}
\usepackage{dcolumn}
\usepackage{bm}
\newcommand{\be}{\begin{eqnarray}}
\newcommand{\en}{\end{eqnarray}}
\newcommand{\ben}{\begin{eqnarray*}}
\newcommand{\enn}{\end{eqnarray*}}

\newcommand{\f}{\frac}

\newcommand{\bi}{\begin{itemize}}
\newcommand{\ei}{\end{itemize}}

\begin{document}
\title{Effects of non-denumerable fixed points in finite dynamical systems}
\author{Sagar Chakraborty}
\email{sagar@bose.res.in}
\affiliation{S.N. Bose National Centre for Basic Sciences, Saltlake, Kolkata 700098, India}
\author{J.K. Bhattacharjee}
\email{tpjkb@iacs.res.in}
\affiliation{Indian Association for Cultivation Of Science, Department of theoretical physics, 2A\&2B Raja S C Mullick Road, Kolkata 700032, India}
\date{April 24, 2007}
\begin{abstract}
The motion of a spinning football brings forth the possible existence of a whole class of finite dynamical systems where there may be non-denumerably infinite number of fixed points.
They defy the very traditional meaning of the fixed point that a point on the fixed point in the phase space should remain there forever, for, a fixed point can evolve as well!
Under such considerations one can argue that a free-kicked football should be non-chaotic.
\end{abstract}
%
\pacs{05.45.-a,47.52.+j,01.80.+b}
\maketitle
\section{Introduction}
Any elementary standard textbook\cite{1} on non-linear dynamics would show that for a finite dynamical system, may be non-linear, one always gets a set of finite number of fixed points.
Interestingly, a seemingly overlooked case in the existing literature had been the one in which the parameters in such a system could depend on the initial conditions; some very recent highly rigourous mathematical works (see \cite{KY} and the references therein) on systems that contain a hidden parameter and thus, continuous one-parameter families of equilibria, have been done. 
As we shall show in this paper, such a finite dynamical system is of real interest, for, it shows an array of weird properties such as evolution of the fixed points, existence set of non-denumerable fixed points etc.
Moreover, as we shall see, such systems are not just of non-realistic pedagogical and academic interest but these systems can in fact exist in reality {\it e.g.,} in the case of a sphere projected with a spin {\it i.e.,} say a football or a tennis ball hit to impart both the linear and angular velocities on them.\\
As we know that the archy path of a kicked football is non-chaotic.
The model equation used to describe it, as we shall see, is in fact such a dynamical system.
As an application of these types of dynamical systems, it will be shown using numerics that the very existence of a {\it bounded} set of non-denumerably infinite fixed points arising in the case of the model equation causes the path of the football to be predictable or non-chaotic.
Thus, the natural existence of peculiar dynamical systems with parameters depending on initial conditions should encourage researchers to explore such systems both theoretically and numerically.\\
In the section (II), we shall recall the true reason for the bending of a ball during its flight and show how the wrong reason (Bernoulli's principle) gives a quantitative estimation of the bending using which we frame the model equation in section (III) that may even model the effect of the reverse Magnus effect.
The model equation is obviously a set of three 1st order non-linear coupled differential equations with a parameter depending on initial conditions, an issue which is the main point being studied in this paper.
In section (IV), we study the model equation analytically and numerically to show that the football's motion is non-chaotic owing to certain peculiarities of the model equation.
The issue of striking peculiarities in the model equation is further taken up in the section (V).
\section{Why football swerves?}
We all have seen expert footballers making weird swerving free-kicks by imparting spin on the football, little knowing, even qualitatively, how exactly it occurs.
Though the bending of the path of a rotating sphere moving through a fluid ({\it e.g.} air) is often attributed to the so-called Magnus-effect\cite{2} (or Robins-Magnus effect\cite{3}, to be historically unerring) owing to the Bernoulli's principle, the process causing the bending of the spinning ball is very much complex even qualitatively, let alone the dream of an exact quantitative equation.\\
%
\begin{figure}
\begin{center}
\includegraphics[width=9cm]{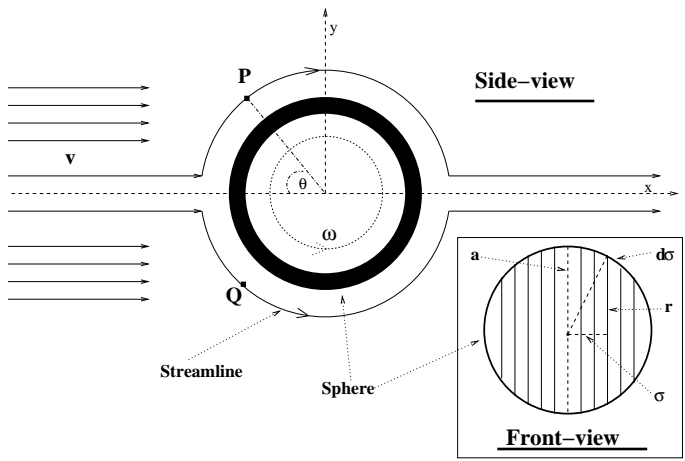}
\end{center}
\caption[{}]{{\bf{Flow past a rotating sphere.}} Here, side-view of an infinitesimal thick disk or radius $r$ has been shown with the adjacent streamlines (shown exaggeratedly away from the body). The front-view shows the necessary geometry needed to carry out the integration.}
\end{figure}
How does Bernoulli's principle\cite{11,12} explain the bending of a rotating ball?
To clarify how exactly the logic goes is given below with the help of an anti-clock-wisely rotating sphere of radius $a$ moving forward in the direction with uniform velocity $v$ transverse to the direction of the axis of rotation, which passes through the center of the sphere, in a fluid (say, air of density $\rho$) at rest. (We call it 1-D case, for obvious reason.)
The angular velocity w.r.t to the center be $\omega$.
Of course if one fixes oneself at the center of the sphere, then in that frame the sphere is rotating and the air is sweeping past the ball uniformly with velocity $v$ (Fig. 1).
The fluid at point P on the streamline adjacent to the periphery of a sliced disc (of radius $r$ and infinitesimal thickness $d\sigma$) of the sphere is supposed to have velocity $v$ of the air in the x-direction and $-\omega r \sin\theta$ and $-\omega r \cos\theta$ in the x-direction and y-direction respectively due to the rotation of the sphere.
The application of Bernoulli's principle suggests that pressure $p$ at point $P$ is:
\be
p=\rho\omega vr\sin\theta-\f{1}{2}\rho\omega^2r^2
\en
The total force $F$ (Magnus force) on the sphere in the -y-direction may be obtained by integrating $p(rd\theta\sin\theta)(d\sigma)$ over the upper and the lower streamlines (remembering $r=\sqrt(a^2-\sigma^2)$):
\be
F=\int_{\sigma=-a}^{\sigma=+a}\int_{\theta=0}^{\theta=2\pi} pr\sin\theta d\theta d\sigma=\f{4}{3}\pi a^3\rho\omega v
\en
Although seemingly okay, any careful reader must have noted that there are flaws in the argument!
Bernoulli's principle, in the form used here, is applicable for the steady, incompressible and inviscid fluid; the first two restriction can be justified but the last one can't be justified near the boundary.
One may note, the no-slip condition has been implicitly assumed in the derivation by allowing the fluid adjacent to the boundary to move with the sphere but this is in direct opposition to the concept of ideal fluid in which Bernoulli's principle is applicable!
Also, Bernoulli's principle can't give the reason for the yet another anomalous effect known as reverse Magnus effect\cite{5}.
The right qualitative argument comes from the consideration of the boundary layer theory due to Ludwig Prandtl\cite{6} who hypothesized that for small viscosity, the viscous forces are negligible everywhere except close to the boundary where the no-slip condition has to be satisfied.
The thickness of the boundary layer approaches zero as the viscosity goes to zero.
As the phenomenon of separation occurs on the line whose points are the singularities of the solutions of Prandtl's equations, the boundary layer separates off the surface owing to the flow against the adverse pressure gradient and behind the sphere, there's the formation of the wake\cite{7,8}, which is not axisymmetric for the rotating sphere; in this case there is upward displacement of the wake and ergo, a downward force on the sphere.
\section{The model equation}
Anyway, it is no way possible to write a general equation for the swerving free-kicked football using the boundary layer theory.
The saving feature is that the actual force is very similar to the Magnus force calculated for the 1-D case.
Consequently, after having explored the true reasons for the deflection, we adopt the Magnus force for the quantitative calculations.
Hence, although a realistic free-kick is a 3-D affair, if the 1-D arguments as discussed in the beginning are applied to this case, we get the model equations.
For that we assume for simplicity\cite{4}, sans the loss of generality:
\begin{enumerate}
\item The off-center kick causes the constant rotation of the football.
\item The shape of the football doesn't get appreciably distorted.
\item Initially, the ball is given no velocity component along y-axis and $\vec{g}$ acts along -z-axis.
\item $\vec{\omega}\bot\vec{v}$ at initial time $t=0$ and $\vec{\omega}$ is in x-z plane making an angle $\alpha$ with the z-axis.
\end{enumerate}
The first assumption needs a bit of justification that has been taken up in the appendix.
By picking the right form for the drag force using empirical $C_D$ vs $Re$ curve for a sphere moving in a fluid at rest, one may write for $10^3\le Re\le10^5$ (which is typical of a free-kick)\cite{14}:
\be
M\dot{\vec{v}}=-k|\vec{v}|\vec{v}-M\vec{g}+\f{4}{3}\pi a^3\rho(\vec{\omega}\times\vec{v})
\label{3}
\en
where $k$ is a constant and equation is subject to the initial condition at $t=0,{}(v_x,v_y,v_z)=(v_{x0},v_{y0},v_{z0})$.
In component form this can be broken down and rearranged as:
\be
\dot{v_x}&=&-A\sqrt(v_x^2+v_y^2+v_z^2)v_x-v_yC\cos\alpha\nonumber\\
\dot{v_y}&=&-A\sqrt(v_x^2+v_y^2+v_z^2)v_y+v_xC\cos\alpha+v_zC\sin\alpha\nonumber\\
\dot{v_z}&=&-A\sqrt(v_x^2+v_y^2+v_z^2)v_z-v_yC\sin\alpha-g
\label{football}
\en
where $A,C,\alpha=\tan^{-1}(v_{z0}/v_{x0})$ are the constant parameters.\\
So, one has to use the set of equations (\ref{football}) as the model equation and fix $A$ and $C$ empirically which can even take care of reverse Magnus effect that can be effected by setting $C$ as negative.
The value of $C$ obtained using Bernoulli's equation in within the order of magnitude of what is achieved using experiments\cite{10}; so we shall stick to that for the sake of further discussion. We choose $A=0.027$ and $C=0.16$ (using Table 1).
%
\begin{center}
\begin{tabular}{|l|l|}
\hline\hline
\multicolumn{2}{|l|}{\bf{Table 1: Values of constants (in SI units)}}\\
\hline\hline
{Circumference of football ($2\pi a$)}&{ 0.70}\\\hline
{ Mass of football ($M$)}&{ 0.42}\\\hline
{ Density of air ($\rho$)}&{1.2}\\\hline
{ Kinematic viscosity of air ($\nu$)}&{ 14.6$\times10^{-6}$}\\\hline
{ Typical velocity ($v$)}&{ 20}\\\hline
{ Typical length scale ($2a$)}&{ 0.22}\\\hline
{ Reynolds number ($Re=(v)(2a)/\nu)$)}&{ $\sim10^{5}$}\\\hline
{ Drag coefficient ($C_D$ for $10^{3}\le Re\le10^{5}$)}&{ $\sim$0.5}\\\hline
{ Number of rotations on the ball ($\omega/2\pi$)}&{ 2}\\\hline
{ Acceleration due to gravity ($g$)}&{ 9.8}\\\hline
{ $A$ of equation (\ref{football}) ($A=0.5\rho C_D\pi a^2/M$)}&{ 0.027}\\\hline
{$C$ of equation (\ref{football}) ($C=1.33\pi a^3\rho\omega/M$)}&{ 0.16}\\
\hline\hline
\end{tabular}
\end{center}
\section{The freekick is non-chaotic}
\begin{figure}[h]
\begin{center}
\epsfxsize=9cm \epsfysize=9cm
\rotatebox{0}{\epsfbox{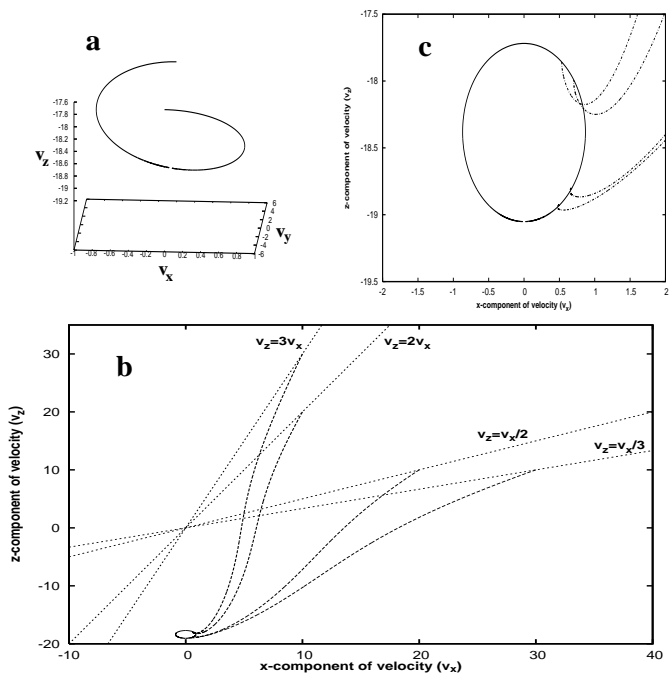}}
\end{center}
\caption[{}]{{\bf{Trajectories and fixed points:}} Going clockwise, Fig. 2a shows the locus of fixed points in 3-D, Fig. 2b (only $v_x-v_z$ plane has been shown for clarity) is showing the collapse of various trajectories (dashed curves, starting form various planes of the form $v_z=cv_x$ marked in the figure) on the locus of fixed points (solid loop), and Fig. 2c is the blown up image of Fig. 2b near the locus of fixed points.}
\end{figure}
\begin{figure}[h]
\begin{center}
\epsfxsize=6cm \epsfysize=9cm
\rotatebox{270}{\epsfbox{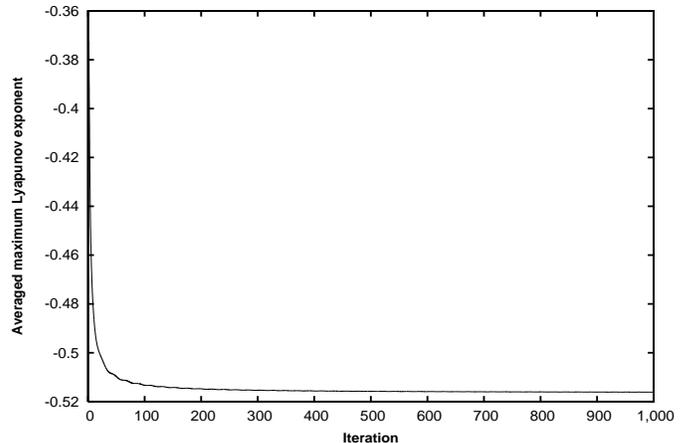}}
\end{center}
\caption[{}]{{\bf{The negative Lyapunov exponent:}} The y-coordinate any arbitrary point on the curve is showing the average of the Lyapunov exponents calculated after every iteration upto that point of the program\cite{13}. Of course, the Lyapunov exponent is settling for a negative value and hence signifying that there should not be any chaos.}
\end{figure}

Now, let's analyze the equations (\ref{football}) purely as a set of non-linear dissipative (and hence, can neither have any quasi-periodic solution nor can have any totally repelling fixed point or orbit) coupled 3-dimensional differential equations. We set $\dot{\vec{v}}=0$ and solve the R.H.S to get following fixed points:
\be
v_x&=&v_x^*\equiv-\Delta^{-1}\{gC^2\sin\alpha\cos\alpha\}\nonumber\\
v_y&=&v_y^*\equiv-\Delta^{-1}\{gHC\sin\alpha\}\nonumber\\
v_z&=&v_z^*\equiv+\Delta^{-1}\{g(H^2+C^2\cos^2\alpha)\}
\en
where, $\Delta=H\{H^2+C^2\}$ and $H$ is the real negative solution of the equation $H^6+C^2H^4-A^2g^2H^2-A^2g^2C^2\cos^2\alpha=0$ which guarantees one value of $H$ for each $\alpha$.
These are either spiral nodes or nodes for all $\alpha$ as can be obtained from the linear stability analysis (and extending the conclusions to this set of non-linear equations, thanks to Hartman-Grobman theorem\cite{9}) which leads to following characteristic equation ($\lambda$ being the eigenvalues and $\xi=\sqrt({v_x^*}^2+{v_y^*}^2+{v_z^*}^2)$):
\be
&&\lambda^3+4A\xi\lambda^2+(C^2+5A^2\xi^2)\lambda+2A^3\xi^3\nonumber\\
&&+AC^2\xi+\f{AC^2}{\xi}({v_x^*}\sin\alpha-{v_z^*}\cos\alpha)^2=0
\en
But note in the definition of $(v_x^*,v_y^*,v_z^*)$ there is $\alpha$ which being $\tan^{-1}(v_{z0}/v_{x0})$ is a function of initial points and so, mathematically, for all $\alpha\in[0,2\pi]$, we have a bounded locus of infinite number of fixed points (Fig. 2a) which being stable forces trajectories to converge on it at different points depending on which plane passing through the y-axis does the initial point of the corresponding trajectory lie (Fig. 2b).
So, any initial point, however far, will evolve in time and collapse on the locus of fixed point, showing that the trajectories can't diverge, thanks to the bounded nature of the finite locus of fixed points.
Hence, we may say that there is no chaos in the system of equations (\ref{football}), for there's no exponential divergence of the trajectories.
This may be quantitatively justified by calculating, of course numerically, the largest Lyapunov exponent of the system of equations (\ref{football}) and finding that the exponent is negative (Fig. 3), thereby confirming that there surely is no chaos in the velocity space of the football.
One may ask, what happens if the ball slows down to below $Re\sim10^3$?
Can chaos come into picture?
The answer probably is in negative since that is basically the Stoke's law (with the possible Oseen's correction) dominated regime where the drag along each direction is linear in the velocity's component in that direction and of course, a set linear equations can't give rise to chaos.
What remains unexamined is what happens precisely when $Re\ge10^5$ when the peculiar effect such as drag crisis\cite{7} and the compressibility effect are switched on.
So what has been arrived at is in accordance with what we see in practice.
A good free-kick specialist who has done enough homework can always almost send the ball where he wants to because the trajectory of the ball is grossly insensitive to the minor changes, natural even for the best of the players, in the initial velocities imparted to the ball.
\section{Peculiarities of model equation}
\begin{figure}[h!]
\begin{center}
\epsfxsize=9cm \epsfysize=9cm
\rotatebox{0}{\epsfbox{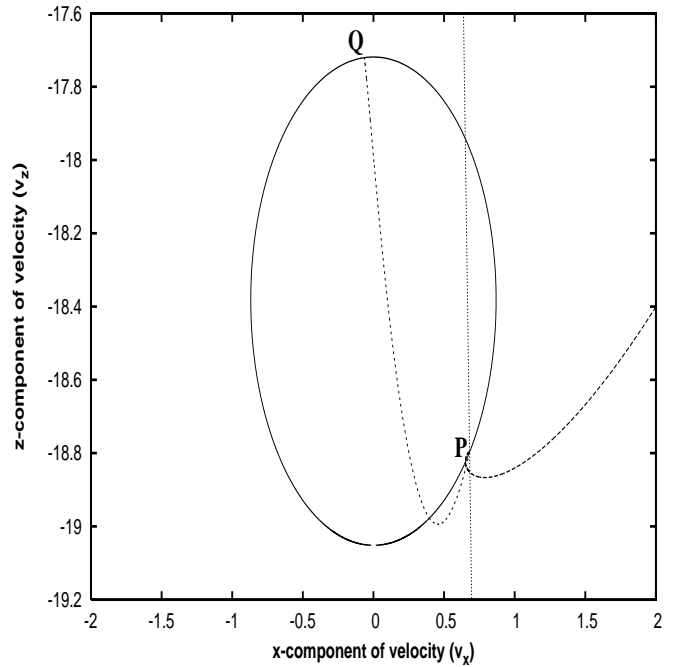}}
\end{center}
\caption[{}]{{\bf{The peculiarity of the fixed points:}} The solid closed loop is the projection of the locus of fixed points on $v_x-v_z$ plane. The curve on the right of point P is the trajectory for the evolution of initial point $v_x=20,$ $v_z=10$ which has reached the loop at the point P that itself lies on the plane $v_z=(-18.8v_x)/0.681$ (straight line passing through the point P) and hence, if is treated as an initial point, evolves as the broken curve on the left of point P to another point Q on the locus of fixed points.}
\end{figure}
Surprises continue to pour in for this simple system. Physically, a fixed point in the phase space means that if one starts from there, one remains there\cite{1}.
One more glimpse at the set of equations (\ref{football}), and one can see that this long held intuitive idea for the fixed point has to be dispensed with at least for the dynamical systems for which there is a parameter depending on the initial conditions themselves.
Any trajectory starting at a point lying on a plane $v_z=cv_x$ (say) reaches a point on the locus of fixed points.
Now, if one starts with the point on the locus of fixed points, the point is bound to be on some other plane $v_z=c'v_x$ (say) and hence should evolve with time to reach another point on the locus (Fig. 4).
So, as time tends to infinity a fixed point on the locus doesn't remain where it has been at $t=0$, rather it moves to some other fixed point and hence in this case, it is the locus of fixed points which maintains itself intact with time rather than the usually believed isolated fixed points.
\section{Conclusion}
To conclude it may be highlighted that much to the delight of mathematicians and others, it has been found in this brief paper that there exists, in the down-to-earth situation of ball-games, a class of dynamical systems (which, surprisingly, has been reported rather recently in the more-than-a-century-old literature on non-linear dynamics which pervades many fields viz. mathematics, physics, chemistry, biology, economics {\it etc.}) where at least one of the parameters is dependent on initial points giving rise to the possibility that there may be non-denumerably infinite number of fixed points.
In the system discussed, the entire phase space got partitioned into a set of infinite number of basins of attraction.
Also, the conventional intuitive definition of fixed point has been shattered because it has been seen here that an initial condition on the fixed point can evolve as well on its own with time.
Besides, it has been shown that a free-kicked football, which basically is such a dynamical system, cannot have chaos during its flight and hence the footballers may take free-kicks without bothering, too much precisely, about the initial velocity he should deliver to the ball.
This is in keeping with what happens in reality.
\begin{acknowledgments}
We thank Debabrata Dutta for helping with the numerics and CSIR, India for providing fellowship to SC, one of the authors, during the period of the research. Also, Mr. Ayan Paul is gratefully acknowledged for providing the authors with relevant scholarly articles.
\end{acknowledgments}
\appendix
\section{CONSTANCY OF THE ANGULAR VELOCITY}
Lets ponder upon the issue of constancy of the angular velocity.
One might argue that the viscosity of the fluid might slow down the rotation of the football (or any other rotating sphere, to be general) as the time passes by and hence taking the angular velocity to be constant in the equation (\ref{3}) is not justified.
True, one knows that there is something called Kirchoff's law in fluids which gives the following expression\cite{7} for the instantaneous torque that slows a rotating sphere of radius $a$ down:
\be
\tau=-8\pi\eta a^3 \Omega
\label{kl}
\en
where, $\eta$ is the dynamic viscosity and $\Omega$ is the instantaneous angular velocity of the sphere.
Obviously, if we denote the moment of inertia of the sphere by the symbol $I$, the torque-balance equation
\be
I\dot{\Omega}=\tau
\label{tb}
\en
would yield the result that the angular velocity decays down exponentially fast.
But, this conclusion hinges on the fact that the relation (\ref{kl}) requires the Reynolds number ($\Omega a^2/\nu$) to be much smaller than unity -- a condition grossly violated in the case being considered in this paper.
Using the table-1, we can see that
\be
\f{\Omega a^2}{\nu}\approx\f{2(2\pi)\times 0.11^2}{14.6\times 10^{-6}}\sim 10^4
\en
where we have taken initial value of $\Omega$ {\it i.e.,} $\omega$.
Therefore, certainly relation (\ref{kl}) is not valid for the case being considered!
\\
So, what is the story here? For finding the analogous expression for $\tau$ in the case of high Reynolds number, lets resort to the powerful and yet simple method of dimensional analysis.
The relevant parameters are $\rho$, $a$ and $\Omega$.
Simple dimensional analysis would then give
\be
\tau=-N\rho a^5\Omega^2
\label{tau2}
\en
Here, $N$ is a numerical constant and the negative sign emphasises that this torque is trying to slow down the sphere's rotation.
This expression might be given the physical interpretation of reaction on the sphere by the fluid due to the transfer of the angular momentum to the fluid by the sphere by dint of its rotation.
Putting relation (\ref{tau2}) in the differential equation (\ref{tb}) and solving with the initial condition: at $t=0$, $\Omega=\omega$, one gets:
\be
\Omega=\f{1}{\f{N\rho a^5t}{I}+\f{1}{\omega}}
\label{Omega}
\en 
Now, with some elementary knowledge in the non-linear dynamics, one knows that a fixed point is reached exponentially fast; relation (\ref{Omega}) suggests that the velocity of the rotational degree of freedom ({\it i.e.}, the angular velocity) slows down only as the inverse of time.
Hence, one may conclude that as far as the behaviour of the trajectories, governed by the equations (\ref{football}), to reach the fixed point is concerned, one can safely choose the angular velocity to be constant.


\begin{thebibliography}{99}
\bibitem{1} S.H. Strogatz, {\it Nonlinear Dynamics And Chaos}, (Perseus Books, Massachusetts, U.S.A) 1994
\bibitem{KY}L.G. Kurakin and V.I. Yudovich, {Chaos {\bf{11}}}, 780(2001)
\bibitem{2} G. Magnus, {Condor's Annalen Dee Phys u. Chemie \bf{8}}, 1(1853)
\bibitem{3} B. Robins, {\it New Principles of Gunnery}, (1742)
\bibitem{4} M. J. Carr$\acute{\textrm{e}}$ , T. Asai, T. Akatsuka and S.J. Hakke, {Sports Engineering {\bf 5}}, 193(2002)
\bibitem{5} R. Mehta, {Annual Review of F1uid Mechanics {\bf 17}}, 151(1985)
\bibitem{6} L. Prandtl, {\it Verhandlungen des dritten internationalen Mathematiker-Kongresses (Heidelburg 1904) Leipzig.}, 1904, pp. 484-491
\bibitem{7} L.D. Landau and E.M. Lifshitz, {\it Fluid Mechanics, 2nd ed.}, (Permagon Press, U.K.) 1987
\bibitem{8} G.K. Batchelor, {\it An Introduction To Fluid Dynamics}, (Cambridge University Press, U.K.) 1967
\bibitem{9} S. Wiggins, {\it Introduction To Applied Nonlinear Dynamical Systems And Chaos}, (Springer, New York, U.S.A.) 1990
\bibitem{10} I. Griffiths, C. Evans and N. Griffiths, {Meas. Sci. Technol {\bf 16}}, 2056(2005)
\bibitem{11} P.K. Kundu and I.M. Cohen, {\it Fluid Mechanics, 2nd ed.}, (Academic Press, Elsevier Science, U.S.A.) 2002
\bibitem{12} D.J. Tritton, {\it Physical Fluid Dynamics 2nd ed.}, (Oxford Science Publications, New York, U.S.A.) 1988
\bibitem{13} J.C. Sprott, {\it Chaos and Time-Series Analysis}, (Oxford University Press, U.S.A.) 2003.
\bibitem{14} B.G. Cook and J.E. Goff, {Eur. J. Phys. {\bf 27}}, 865(2006)
\end{thebibliography}
\end{document}